\documentclass[published]{JHEP}
\usepackage{epsfig}

\newcommand\fverb{\setbox\pippobox=\hbox\bgroup\verb}
\newcommand\fverbdo{\egroup\medskip\noindent%
			\fbox{\unhbox\pippobox}\ }
\newcommand\fverbit{\egroup\item[\fbox{\unhbox\pippobox}]}
\newbox\pippobox
\newcommand{\beq}{\begin{equation}}
\newcommand{\eeq}{\end{equation}}
\newcommand{\beqa}{\begin{eqnarray}}
\newcommand{\eeqa}{\end{eqnarray}}
{\newcommand{\gsim}{\mbox{\raisebox{-.6ex}{~$\stackrel{>}{\sim}$~}}}
{\newcommand{\lsim}{\mbox{\raisebox{-.6ex}{~$\stackrel{<}{\sim}$~}}}
\def\sfrac#1#2{{\textstyle{#1\over#2}}}

\def\sss{\scriptscriptstyle}
\def\sQ{{\sss Q}}
\def\Oeff{\Omega_\Lambda^{\rm eff}}

\title{Quintessence, Cosmological Horizons, and Self-Tuning}

\author{by J.M. Cline\\
McGill University, 3600 University St., Montr\'eal, Qc H3A 2T8, Canada\\
	E-mail: \email{jcline@physics.mcgill.ca}}

\preprint{\hepth{9912999}}	

\abstract{We point out that quintessence with an exponential potential 
$V_0 e^{-\beta Q/\sqrt{3}M_p}$ can account for the present observed
acceleration of the universe, without necessarily leading to eternal
acceleration.  This occurs for $2.4 < \beta < 2.8$.  Thus a cosmological
horizon, which is supposed to be problematic within the context of string
theory, can be avoided.  We argue that this class of models is not
particularly fine-tuned.  We further examine this question in the context
of a modified  Friedmann equation, $H^2 \propto \rho + p$, which is
suggested by higher dimensional self-tuning approaches to the cosmological
constant problem.  It is shown that the self-tuning case can also be
consistent with observations, if $1.8 < \beta < 2.4$.  Future observations
of high-$z$ supernovae will be able to test whether $\beta$ lies in the
desired range.}

\begin{document} 

\maketitle 

\section{Introduction}

There is convincing evidence that the universe is presently dominated by a
form of dark energy density which is decreasing significantly more slowly
with time than the energy density of ordinary matter.  The first Doppler
peak in the cosmic microwave background fluctuations is strongly consistent
with a flat universe \cite{cmb}, whose density is the critical one
$\rho_c$, yet the matter energy density is known by other means to be no
more than $\Omega_m = \rho_m/\rho_c \sim 0.3$ \cite{Om}.  Recent
observations of the higher peaks by the various CMB experiments give
$\Omega_m = 0.25$ \cite{smoot}.  The Hubble diagram deduced from high
redshift type I supernovae provides independent evidence for an additional
component $\Omega_\Lambda$ of the energy density \cite{SN}.   From an
empirical viewpoint, pure vacuum energy ($\Lambda$) is the simplest
explanation, but theoretically it is difficult to explain why $\Lambda$ is
some 124 orders of magnitude smaller than the natural scale set by the
Planck mass, $M_p^4$.  If the dark energy is due to a rolling scalar field
$Q$, quintessence \cite{quint}, whose potential energy vanishes as
$Q\to\infty$, this might be a more natural explanation for why the dark
energy density is small.

A further motivation for quintessence could be coming from string theory,
because of its apparent incompatibility with de Sitter space \cite{banks}.  
An eternally accelerating universe, which would result from a positive
cosmological constant, seems to be at odds with string theory, because of
the impossibility of formulating the S-matrix.  In de Sitter space the
presence of an event horizon, signifying causally disconnected regions of
space, implies the absence of asymptotic particle states which are needed
to define transition amplitudes.  Quintessence, on the other hand, would
seem to offer the possibility of temporary acceleration to account for
current observations, without necessarily making the scale factor of the
universe accelerate forever.

Recent papers have pointed out that, in fact, quintessence generically
does lead to eternal acceleration, if the universe is accelerating now
\cite{Paban,HKS} (see also \cite{He}). Therefore it is concluded that
string theory is equally at odds with quintessence or a positive
cosmological constant.  The purpose of the present paper is to explore
some simple loopholes to this conclusion.\footnote{Ref.\ \cite{moffat}
has also addressed this question in the context of gravitiy with a
time-varying speed of light.}

Our initial motivation was an interesting possibility coming from the
brane-world scenario, in particular, attempts to address the cosmological
constant problem through self-tuning solutions to the Einstein equations
\cite{st}. In this approach, it is assumed that our universe is a 3-brane
with arbitrary tension $\Lambda$, embedded in an extra dimension.  A
scalar field living in the extra dimension adjusts itself so as to yield a
static solution to the Einstein equations, regardless of the value of
$\Lambda$, which otherwise would act like the 4-D cosmological constant
and lead to inflation of the brane.  Although there are many problems with
this idea \cite{stp}, ref.\ \cite{CM} explored the question of how
cosmology would be affected for brane observers assuming an acceptable
model of self-tuning was found.  If the scalar couples only to the volume
element $\sqrt{g}d^{\,4}x$ of the three-brane, \cite{CM} showed that it is
possible to obtain a modified Friedmann equation of the form
\beq
\label{MFE}
	H^2  = \left({\dot a\over a}\right)^2 = 2\pi G(\rho + p)
\eeq
plus corrections of order $G(\rho+p)^2/$TeV$^4$ (assuming the 5-D quantum
gravity scale is of order TeV).  This is a very interesting twist on
normal cosmology because (1) it leads to no expansion in the case of
vacuum energy, where $p=-\rho$; (2) it is indistinguishable from the
normal equation during the radiation dominated era, since $p=\rho/3$;
and (3) it gives a Hubble rate only $\sqrt{3/4} = 0.87$ times smaller
than normal during the matter dominated era, which would be difficult to
distinguish from the standard value given the uncertainties on $\Omega_m$
(the fraction of the critical density in matter) and $H_0$ (the present
value of the Hubble parameter).  Since the modified Friedmann equation
eliminates conventional inflation, it would remove the obstacle to
defining an S-matrix in the 4-D universe when $\Lambda >0$.  We might
also expect it to ameliorate the problem with eternal acceleration in
quintessence models.  

Moreover, if the equation of state for the dark energy, $w = p/\rho$, turns
out to be $w>-1$, as is still allowed by the supernova data, this could be
indirect evidence for a modified Friedmann equation.  Taking the time
derivative of (\ref{MFE}), one can show that the acceleration is
\beq
\label{accel}
	{\ddot a\over a} = -\pi G\rho (1+w)(1+3w)
\eeq
so that the condition for positive acceleration is $-1 < w < -1/3$. This is
to be contrasted to the standard result, ${\ddot a\over a} = -(4\pi
G/3)\rho  (1+3w)$, requiring only that $w < -1/3$.  In (\ref{accel}), as
$w$ approaches $-1$, the acceleration would disappear, coming into conflict
with the observations.  It must be emphasized however that since the
connection between $w$ and acceleration is no longer the same when  eq.\
(\ref{MFE}) is adopted, one should the modified expansion law.

In the following we will present results for exponential quintessence
potentials, since these proved to be the most promising for overcoming the
horizon problem.  As will be shown, the modified Friedmann equation
enlarges the range of potential parameters which avoid a future horizon,
but there is also an allowed range using the normal Friedmann equation.
In the penultimate section it will be argued that these solutions do not 
require any more fine tuning than those arising from other models 
of quintessence.

\section{Quintessence evolution and event horizon}

The quintessence field $Q$ with potential $V$ has the usual equation 
of motion, $\ddot Q + 3H\dot Q + {dV\over dQ} = 0.$  Its pressure and
energy density are given by $\rho_\sQ = \sfrac12\dot Q^2 + V$ and
$p_\sQ = \sfrac12\dot Q^2 - V$.
To explore cosmology both in the usual case and with the modified
Friedmann equation (\ref{MFE}) we will introduce a parameter $x=0,1$
and write the Hubble rate as
\beq
	H^2 = \kappa^2_x \left( (1 + \sfrac{x}{3})\rho_r + \rho_m + 
	\sfrac12 (1+x)\dot Q^2  + (1-x) V \right)
\eeq
where $\kappa^2_x = 8\pi G/(1+\sfrac{x}{3})$ and $\rho_r$ and $\rho_m$
are the radiation and matter energy densities, respectively.   It thus
reduces to the standard equation when $x=0$ and the self-tuning one
(\ref{MFE}) when $x=1$.  Guided by the CMB data, we assume the curvature
term in $H^2$ to be absent.

Rather than integrating with respect to time $t$, it is convenient to
think in terms of redshift, $1+z = 1/a(t)$, where we take the
present scale factor $a(t_0)$ to be unity.  Further defining $u = \ln(1+z)$,
and rescaling the fields and energy densities via
\beqa
  \hat Q &=& \kappa_x Q, \qquad\qquad\!\! \hat\rho_i = \kappa_x^2
H_0^{-2}\rho_i,
  \nonumber\\
  \hat V &=& \kappa_x^2 H_0^{-2} V, \qquad\hat H = H/H_0,
\eeqa
the quintessence equation of motion and Friedmann equation can be
written in the dimensionless form
\beqa
	\hat Q'' &=& \hat H^{-2}\left[ \left((1 + \sfrac{x}{3})\hat\rho_r +
	\sfrac32\hat\rho_m + (1-x)\hat V\right)\hat Q'
	- \left(1-x\hat Q'^2\right) {\partial\hat V\over \partial\hat Q}
	\right]\\
	\hat H^2 &=& {(1 + \sfrac{x}{3})\hat\rho_r + \hat\rho_m
	+ (1-x)\hat V\over 1 -\sfrac12(1+x)\hat Q'^2},
\eeqa
where primes denote ${d\over du}$.  The matter and radiation densities
scale with $u$ like $\hat \rho_m =\hat\rho_{m,0} e^{3u}$ and  $\hat \rho_r
=\hat\rho_{r,0} e^{4u}$ with respect to their present values at $u=0$. 
These dimensionless equations are well-suited to numerical integration,
which is the main technique of our investigation. Notice that conventional
time $t$ has been eliminated, and $u=\ln(1+z)=-\ln(a)$ now plays the role of
the time variable.

In comparing the properties of the quintessence field to observations,
we will refer to the fractions of the critical energy density, and the
equation of state.  The former are defined as
\beqa
\label{Oieq}	
\Omega_i &=& {\kappa_x^2 \rho_i\over H^2} = {\hat\rho_i\over \hat H^2};\\
\label{Oqeq}
\Omega_\sQ &=& {\kappa_x^2\over H^{2}}\left({1+x\over 2}\,
	\dot Q^2 +(1-x)V\right) = 
	{1+x\over 2}\,\hat Q'^2 +{1\over \hat H^{2}}(1-x)\hat V,
\eeqa
which satisfy $\Omega_r + \Omega_m + \Omega_\sQ = 1.$  
As for the quintessence equation of state, it is given by
\beq
\label{weq}
	w = { \dot Q^2 - 2V \over \dot Q^2 + 2V } = 
{ \hat H^2 \hat Q'^2 - 2\hat V \over \hat H^2 \hat Q'^2 + 2\hat V }
\eeq
This follows from the fact that $\dot Q = - H Q'$.  

In standard cosmology, $w<-1/3$ is the criterion for acceleration.  But as
noted in the introduction, the relation between $w$ and acceleration is
modified for the self-tuning scenario with $x=1$.  It is therefore useful
to have another quantity indicative of acceleration, which can be more
directly related to the observations of high-$z$ supernovae.  Let us first
review what is actually constrained \cite{white}:
 it is the distance modulus ($m-M =$
apparent minus absolute magnitude) of the SN versus its redshift, where
$m-M = 5\log_{10}(d_L/\hbox{Mpc}) + 25$, and the luminosity distance
$d_L(z)$ is given by $(1+z)H_0\Delta(z)$, with
\beq
	\Delta(z) = \int_{t(z)}^{t_0} {dt\over a(t)} = 
\int_0^z {dz'\over\hat H(z')} = \int_0^{\ln(1+z)}
	 {e^u\over \hat H(u)}\, du
\eeq
In a flat universe with only matter and cosmological constant components,
$\Omega_m+\Omega_\Lambda = 1$,
one would have
\beq
\label{DLeq}
 \Delta_\Lambda \equiv \int_1^{1+z} 
	{dx\over (x^3\Omega_m + \Omega_\Lambda)^{1/2}}
\eeq
The high-$z$ SN results essentially try to measure $\Delta(z)$ as a
function of $z$ in order to fit $\Omega_\Lambda$.  We therefore define a
phenomenological parameter, $\Oeff$:
\beq
\label{Oeffeq}
	\Oeff \equiv \Omega_\Lambda\hbox{\ such that\ } \Delta_\Lambda(3/4) = 
	\Delta(3/4),
\eeq
{\it i.e.,} $\Oeff$ is the amount of $\Omega_\Lambda$ for
pure cosmological constant that would be required to give the same value
of $\Delta(z)$ as is produced by the quintessence model.  This is a
$z$-dependent definition, and we somewhat arbitrarily take the value
$z=0.75$ because this is roughly the redshift where the current SN data
are which are the most sensitive to $\Omega_\Lambda$.

The function $\Delta$ is also relevant for determining whether an event
horizon appears in the spacetime, since the coordinate distance traveled
by a photon between the present and a future (negative) value of $z$
is precisely $\Delta(z)$.  The criterion for an event horizon is that
$\lim_{z\to -1}\Delta(z) < \infty$.  In this case the photon travels
a finite coordinate distance in an infinite time, and this determines the
position of the horizon for the observer that emitted the photon: no signals
originating from beyond that position will ever be able to reach him.
For ease of representation, we will define the following measure of
horizon formation:
\beq
\label{Deq}
D \equiv \lim_{a\to\infty} {\partial\ln\Delta\over\partial\ln a}
	= \left\{ \begin{array}{ll} 0, & \quad a(t) \sim e^{H t},\ 
	\hbox{horizon exists} \\
	{1\over q} - 1, & \quad a(t)\sim t^q,\ \hbox{no horizon if\ }q<1
\end{array}
	\right.
\eeq
Here $q={2\over 3(1+w)}$ if the dominant component has equation of state
$p=w\rho$.  For example, $D=1/2$ in a universe which behaves as though 
it is matter dominated ($q=2/3$) at very late times.  We will see below
that $D=1/2$ is the maximum value that arises in the quintessence
models which we consider.

\section{Results}

We have examined some of the popular choices of the quintessence 
potential, $V(Q)$, including inverse powers, exponentials, and combinations
of the two.  For avoiding the event horizon, the exponential potential
(first considered in \cite{PR}, and subsequently in \cite{wett}-\cite{clw}) 
seems most promising:
\beq
	\hat V(\hat Q) = \hat V_0 e^{-\beta\hat Q}
\eeq
This potential has only a single free parameter, $\beta$,\footnote{in the
notation of ref.\ \cite{wett}, $\beta = \sqrt{6}a$, and in that of refs.\
\cite{FJ,clw}, $\beta=\sqrt{3}\lambda$,  assuming the normal Friedmann
equation.}  once the constraint that $\Omega_\sQ = 1 - \Omega_m$ is imposed
for the present epoch, for this determines $V_0$.  The statement is
strictly true if one assumes that initially $\hat Q'=0$, since the initial
value of $\hat Q$ itself can be absorbed into the definition of $\hat
V_0$.  However, even if $\hat Q'\neq 0$ initially, the quintessence field
converges to an attractor solution \cite{wett}--\cite{clw}, as shown in
figure \ref{fig1a}: the two solutions corresponding to different initial
conditions converge to the same functional form after some time.  In this
example, we numerically integrated the equation of motion for the case
$\Omega_m = 0.25$ starting from initial conditions at $u=\ln(1+z) = 12$
($z=1.6\times 10^{5}$), during the radiation dominated era with initial
conditions $\hat Q = 0$\footnote{The initial condition on $\hat Q$ is not
significant since it can be absorbed into $\hat V_0$.} and $\hat Q' = 0$ or
$1$.  Although the early behavior of $\hat Q$ is clearly affected by the
difference in initial $\hat Q'$, both solutions join their common
trajectory well before quintessence starts to dominate in the present
era.   Figure \ref{fig1b}\ shows the corresponding quintessence equation of
state, $w$, in the two cases.  In this example it is clear that
quintessence can contribute to the acceleration during the period when
$w=-1$, whereas $w\to -0.25$ in the future, which is larger than $-1/3$ and
therefore cannot cause acceleration.

\FIGURE{
\centerline{\epsfxsize=3.5in\epsfbox{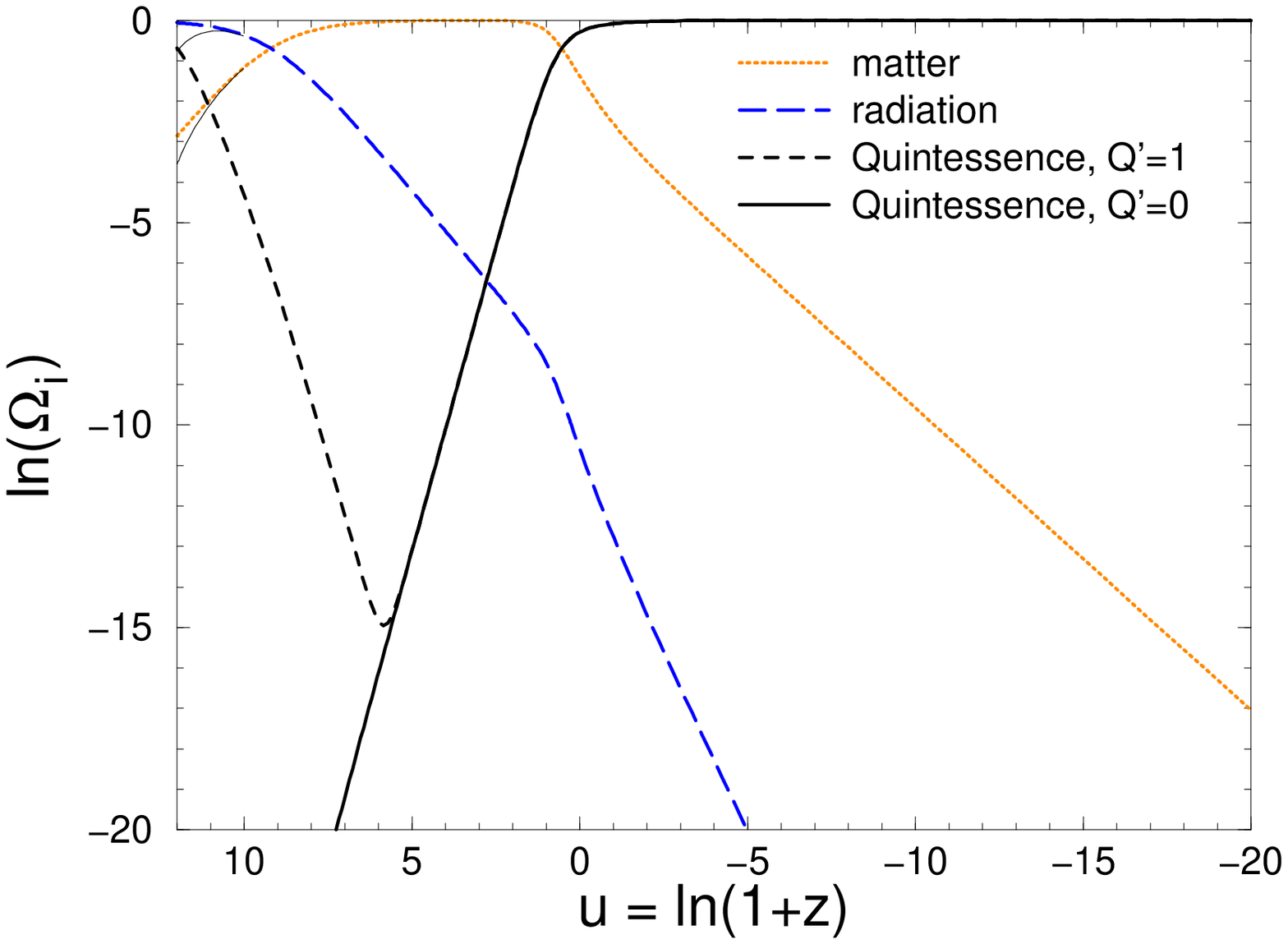}}
\vspace{-1.0cm}
\caption{$\Omega_r$, $\Omega_m$ and $\Omega_\sQ$ as a function of 
$u = \ln(1+z)$ for the potential $\hat V = \hat V_0 e^{-2.6 \hat Q}$
({\it i.e.,} $\beta=2.6$).  The initial conditions at $u=12$ are
$\hat Q'=0$ and $\hat Q'=1$, respectively.}
\label{fig1a}}

\FIGURE{
\centerline{\epsfxsize=3.5in\epsfbox{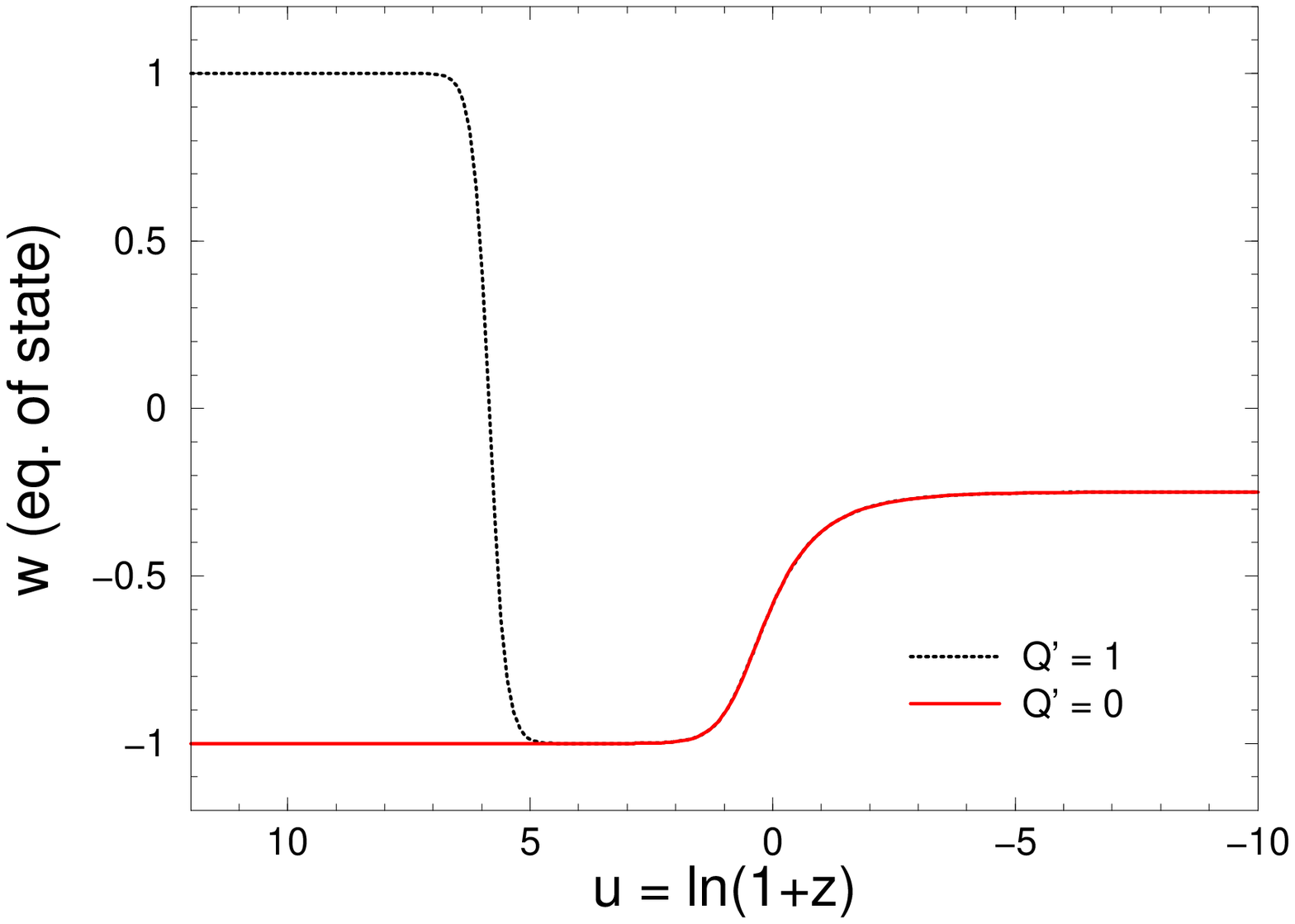}}
\vspace{-1.0cm}
\caption{The quintessence equation
of state, $w$, as a function of $u$, for the same two sets of
initial conditions as in figure \ref{fig1a}.}
\label{fig1b}}

Now let us consider the fate of the universe at very late times, for the
interesting range of potential parameters. We find that a range of $\beta$
values exists such that the cosmological expansion is accelerating today
even though at late times it will revert to a power law, $a\sim t^q$, with
$q\le 2/3$.  This is illustrated by the solid curves of figure \ref{fig2},
which again were made assuming that $\Omega_m=0.25$ and the initial
conditions $\hat Q=\hat Q'=0$ at $u=12$. From the middle curve, the horizon
formation parameter $D$ (eq.\ (\ref{Deq})), we see that an event horizon is
avoided if $\beta \gsim 2.4.$  It is  interesting to note that for larger
values of $\beta$, $D\to 1/2$, just as it would for a matter-dominated
universe, even though quintessence is dominating at late times.  This
happens because the solutions have the  property that $\rho_\sQ \sim
a^{-3}$, just as though the universe was matter-dominated \cite{FJ}.

However, very large values of $\beta$ do not yield sufficient
acceleration at present times to be consistent with observations.  This
can be seen from the top and bottom curves, showing $\Oeff$ (defined in
eq.\ (\ref{Oeffeq})) and $w$ (eq.\ (\ref{weq})) respectively.  The 99\%
confidence level SN limits are $w<-0.5$ for $\Omega_m=0.25$ and $\Oeff >
0.5$ for a flat universe.  It is difficult to apply the limit on $w$ in a
quantitative way in the present model because of the fact that $w$ is
changing very rapidly between the redshifts of $z=1/2$, where much of the
SN data is clustered, and the present, $z=0$.  This rapid variation can be
seen in the example of figure \ref{fig1b}\ as well as by comparing the
bottom sets of curves in figure \ref{fig2}.  On the other hand, the limit
on $\Oeff$ can be applied in a straightforward way to give $\beta \lsim
2.8$.\footnote{Naively one might expect that $\Oeff =1-\Omega_m$ should be
satisfied, which would rule out all values of $\beta$ shown, but we remind
the reader that $\Oeff$ is defined as the amount of real $\Omega_\Lambda$
that would give the same amount of current acceleration as the given
quintessence model.  Since quintessence is less efficient at causing
acceleration than is vacuum energy, it is not surprising that $\Oeff$ is
less than $\Omega_\sQ$.} Thus it is possible to satisfy the observational
constraints without getting an event horizon if $2.4 \lsim \beta \lsim
2.8$.

\FIGURE{
\centerline{\epsfxsize=5.5in\epsfbox{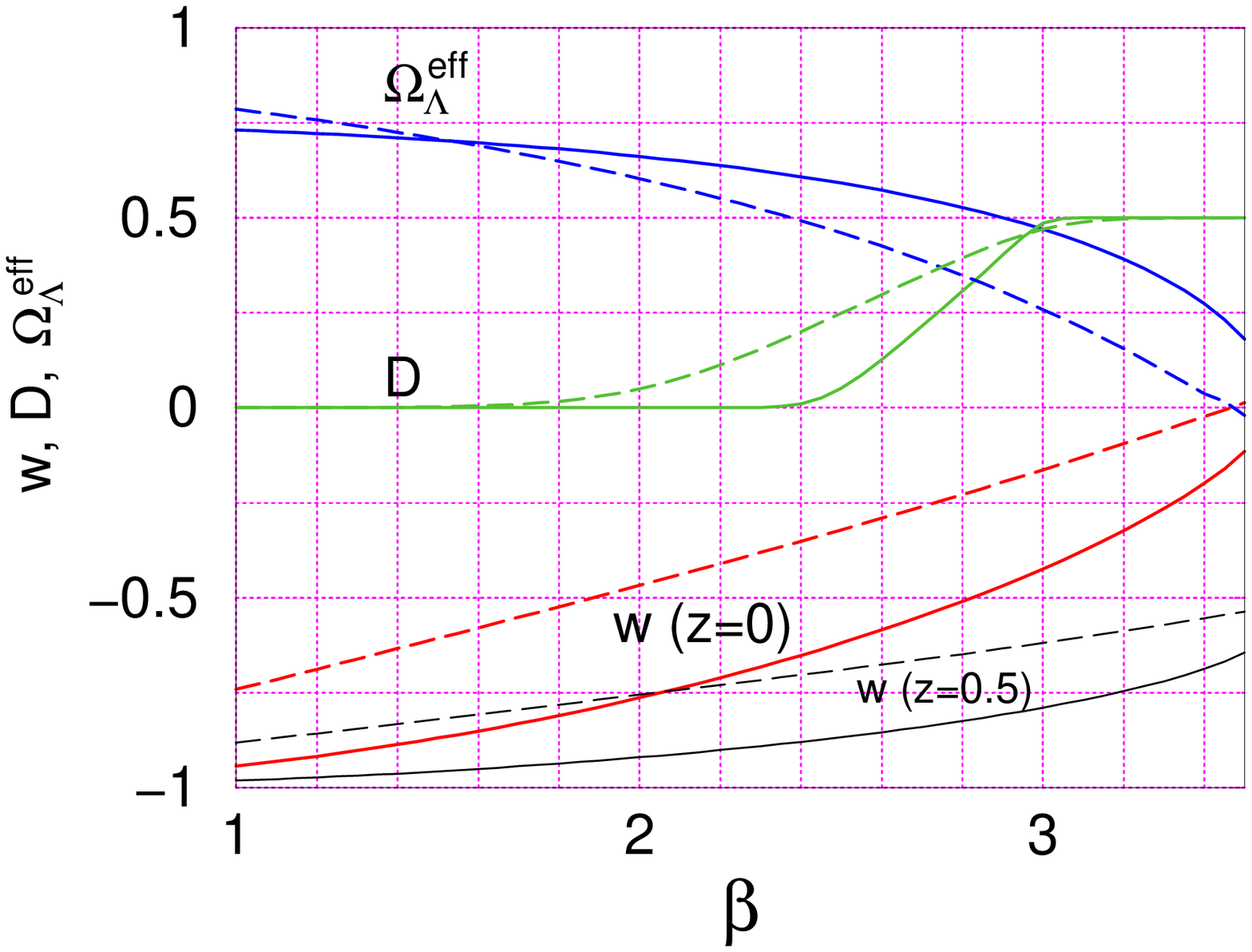}$\phantom{AAAAAA}$}
\caption{From bottom to top: the quintessence equation of 
state $w$, at redshifts $z=1/2$ and at $z=0$, the horizon formation
parameter $D$ (eq.\ (\ref{Deq})), and
the SN effective $\Omega_\Lambda$, $\Oeff$ (eq.\
(\ref{Oeffeq})),
as a function of the potential parameter $\beta$.  Solid lines are for the
normal Friedmann equation ($x=0$), dashed are for the self-tuning one
($x=1$).}
\label{fig2}}

We have also done the same analysis using the self-tuning Friedmann
equation, shown by the dashed curves of figure \ref{fig2}.  One sees that
the range of $\beta$ for which no horizon forms is enlarged to $\beta\gsim
1.8$.  This agrees with the intuitive expectation that acceleration is
reduced in this case, compared to the standard Friedmann equation. But at
the same time, $\Oeff$ and $|w|$ are decreased, making it more difficult
to obtain the observed acceleration.  Since the connection between ${\ddot
a(t)\over a}$ and $w$ is no longer the same as assumed in the analysis of
the SN data, we again take advantage of the $\Oeff$ parameter.  Demanding
that $\Oeff > 0.5$ gives $\beta < 2.4$.  The allowed range consistent with
no horizon, $1.8 \lsim\beta\lsim 2.4$, is thus shifted and slightly
widened relative to the normal case.

It is easy to elucidate the origin of our loophole to the horizon-formation
arguments put forward in references \cite{Paban}-\cite{He}.  These analyses
assume that quintessence is dominating very strongly in the present, or
equivalently that its equation of state does not change from its present
value.  However this approximation is not valid for the solutions we have
presented, such as in figure \ref{fig1a}.  In the present epoch, $u=0$, the
equation of state $w$ for these solutions is always going through a
transition from $-1$ to some value greater than $-1/3$, so as to avoid the
horizon.  In the case of solutions with small $\beta<2.4$, it is also true
that $w$ changes near $u=0$, but its final value at large times is
$w<-1/3$, leading to a horizon.

\section{Naturalness}

In this section we discuss the question of how much fine tuning is needed
to obtain the desired solutions.  The exponential potential has been
somewhat disparaged because of the emphasis on solutions which reach the
scaling regime, where $w$ is constant, very early in the evolution.  Such
solutions are uninteresting in light of the current data  because they
maintain a constant value of $\Omega_\sQ$.  Since $\Omega_\sQ$ must be less
than about 15\% at nucleosynthesis or during large scale structure
formation, this would render its contribution too small to account for the
present acceleration. But as pointed out in \cite{swz}, this negative
conclusion can be circumvented by assuming that quintessence is far from
the late-time attractor solution in the not-too-distant past, so that $w$
can evolve, which is exactly the situation for the solutions presented
here.

Does the fact that these quintessence solutions start out far from the
late-time attractors make them less natural?  We argue that this is not the
case.  All quintessence models require one tuning in order to achieve
$\Omega_\sQ = 1 - \Omega_m$ today, and this is the only one which we
have invoked.  The tuning is imposed as a particular
value of the combination
\beq
	\tilde V_0 \equiv V_0 e^{-\beta\hat Q_i},
\eeq
where $\hat Q_i$ is the initial value of $\hat Q$.  The value of 
$\tilde V_0$
required to make $\Omega_\sQ = 1 - \Omega_m$ today depends on the
initial velocity, or equivalently $\hat Q'_i$.  In the very early universe
the kinetic energy of quintessence typically dominates over its potential
energy unless $\hat Q'_i$ is exactly zero (see eq.\ (\ref{Oqeq})), so this 
amounts to a choice for the initial value of $\Omega_{\sQ,i} \cong
(1+x)\hat Q_i'^2/2$.  Figure \ref{fig1a} shows that whether $\Omega_{\sQ,i}
\cong 0 $ or $1$, the recent evolution of the quintessence is identical, so
long as $\tilde V_0$ takes the right value.  In this sense, we can say that
the models under consideration are very insensitive to the initial
conditions. This conclusion in no way depends on our choice for the initial
time.  Figure \ref{fig4} shows the evolution of $\Omega_i$ for the same
parameters as in fig.\ \ref{fig1a}, except now the initial redshift is
taken to be $10^{17}$, corresponding to an initial temperature of 100 TeV.
Not only is the late-time evolution unaffected, but the choice of $V_0$
is identical if $\hat Q'_i=0$, and $V_0$ only changes by a factor of
$5$ relative to the later initial condition if $\hat Q'_i\neq 0$.
\vspace{-0.25cm}
\FIGURE{
\centerline{\epsfxsize=3.5in\epsfbox{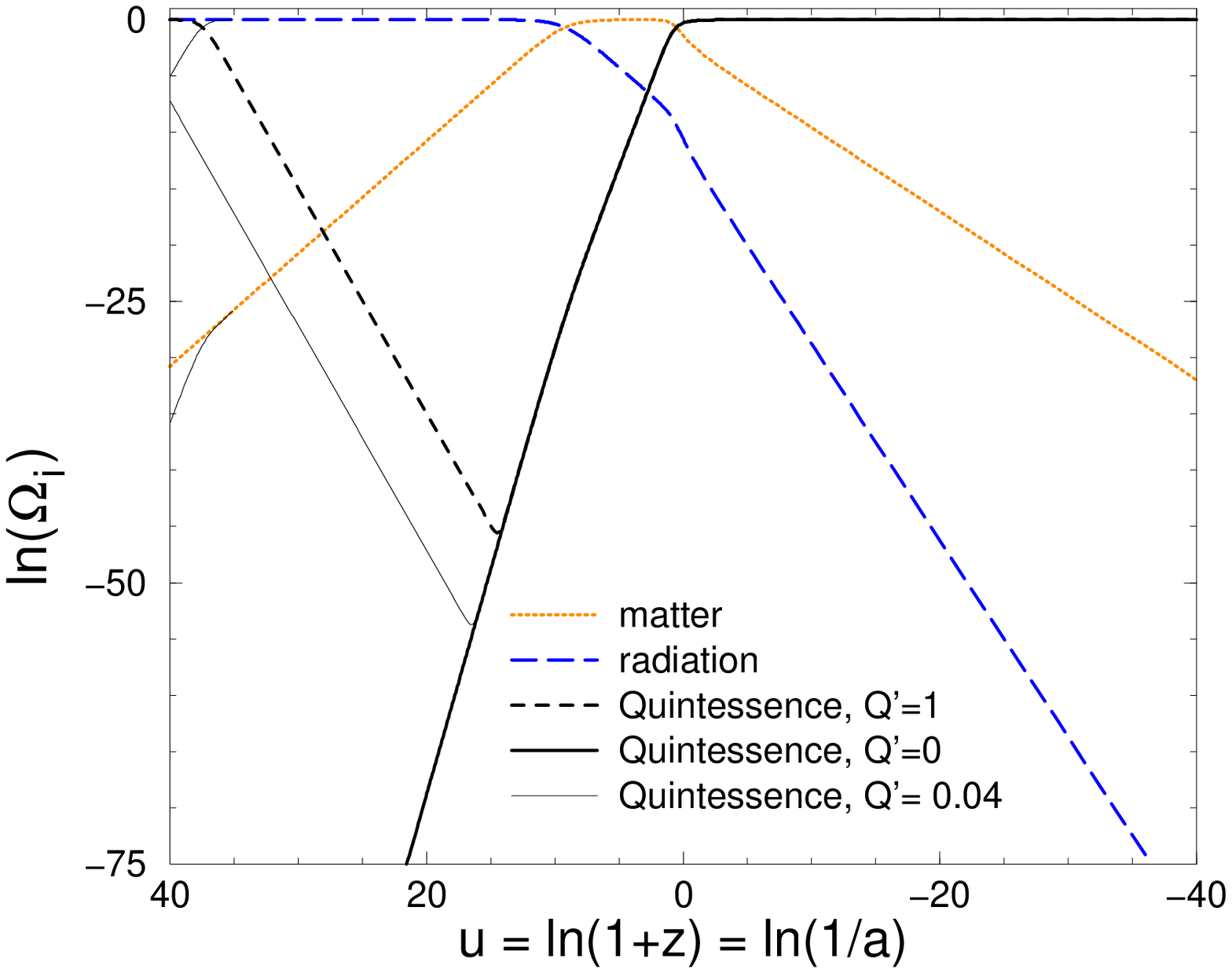}}
\vspace{-1.0cm}
\caption{Same as fig.\ \ref{fig1a}, except starting at redshift
$z=10^{17}$ instead of $z=10^5$ (and the additional initial condition
$\hat Q'_i=0.04$ is also shown, which corresponds to inflationary initial
conditions with equipartition of energy).  
This demonstrates the insensitivity
of the solution to the choice of the initial time.}
\label{fig4}}

Another aspect of naturalness is the value of $\tilde V_0$ required.  If
$\hat Q'_i \ll 1$ (hence $\Omega_{\sQ,i}\ll 1$), as would be natural  in
inflation if equipartition was realized \cite{FJ,swz},  then $\tilde V_0$
must be of order the present critical density,  $\tilde V_0\sim (10^{-3}$
eV$)^4$.  (The evolution with such an initial condition, with
$\Omega_{\sQ,i} = 10^{-3}$, hence $\hat Q'_i = 0.04$, is illustrated in
figure \ref{fig4}.)   This looks unnaturally small in particle physics
units, but one advantage of the exponential potential is that the smallness
of $\tilde V_0$ can be explained by a moderately large value of $\hat
Q_i$.  For example,  if $V_0\sim$ (TeV)$^4$, then $Q_i$ should be of order
$50 M_p$, which is not such a disturbing hierarchy.

The final aspect of naturalness is how sensitive the solution is to small
changes in $V_0$.  Changing $V_0$ causes a shift in the time when the
quintessence and matter energy densities become comparable, which is related
to the so-called coincidence problem: why is it that quintessence is just
starting to dominate in the present epoch?  
It can be shown that scaling the potential by a factor of
$V_{\rm new}/V_0$ leads to the following dependence in the redshift of
matter-quintessence equality, $z_{m-q}$:
\beq
	z_{m-q} = 0.71 + \left(V_{\rm new}\over V_0\right)^{1/3}.
\eeq
The power $1/3$ is just coming from the fact that $V_0$ must be of
order the critical density at $z_{m-q}$, and density scales with redshift
like $(1+z)^3$.  Therefore, in some sense the value of $z_{m-q}$ is rather
insensitive to the value of $V_0$; the coincidence problem would be much
worse if the dependence was through a higher power.

\section{Conclusion}

We have shown that it is possible to evade the cosmological event horizon
which might pose a difficulty for deriving quintessence from string
theory: for a range of $2.4 \lsim \beta \lsim 2.8$ with the conventional
Friedmann equation, or $1.8 \lsim\beta\lsim 2.4$ for the self-tuning
variant, the exponential potential $V(Q) = V_0 e^{-\beta\kappa_x Q}$
gives this outcome.  We also tried other kinds of potentials, such
as inverse powers, $V\sim Q^{-p}$, but for these the development of 
a future horizon was found to be inevitable.

We began this work with the idea that a self-tuning Friedmann equation
might make it easier to avoid a cosmological horizon in a
quintessential universe.  The outcome is that self-tuning does not really
make a big difference: eternal acceleration can be avoided with or without
self-tuning.  Of course self-tuning is still very interesting, because
it allows us to work with a larger class of potentials,
\beq
	V = V_1 + V_0 e^{-\beta\hat Q}
\eeq
since only in the self-tuning case is the evolution completely insensitive
to the value of $V_1$.  It is likely that there are other problems with
self-tuning, since the strength of gravity on subgalactic scales is known
to be consistent with the normal Planck mass, whereas gravity looks 
effectively weaker on cosmological scales in the self-tuning case.  This is
suggestive of the presence of an extra scalar component like a  massive
Brans-Dicke field, whose limited range accounts for the difference between
the effective Planck mass at large and small distance scales.  However the
couplings of such a field to matter are very highly constrained by precision
tests of general relativity in the solar system, like the precession of the
perihelion of Mercury.\footnote{I thank Maxim Pospelov for discussions on
this point.}

Regardless of self-tuning however, the class of solutions we have
discussed  seem sufficiently natural to warrant consideration as a strong
candidate for the dark energy which is presently observed.  It is expected
that  progress in the observations of high-$z$ supernovae (the SNAP 
experiment \cite{snap}) will soon be able to  distinguish  this kind of
model from others through an accurate determination of the time dependence
of the equation of state \cite{snap-theory}.  The very large present
time-dependence of $w$ in the exponential models  makes them particularly
interesting in this respect.  Figure \ref{fig5} shows $dw/dz$ at redshift
$z=0.5$ for the relevant range of $\beta$.  Thus if the SNAP experiment was
to measure that $dw/dz < -0.26$ at $z=0.5$, it would indicate that the
universe will stop accelerating in the future and thus avoid an event
horizon, in the context of the model discussed here.

\vspace{-0.1cm}
\FIGURE{
\centerline{\epsfxsize=3.5in\epsfbox{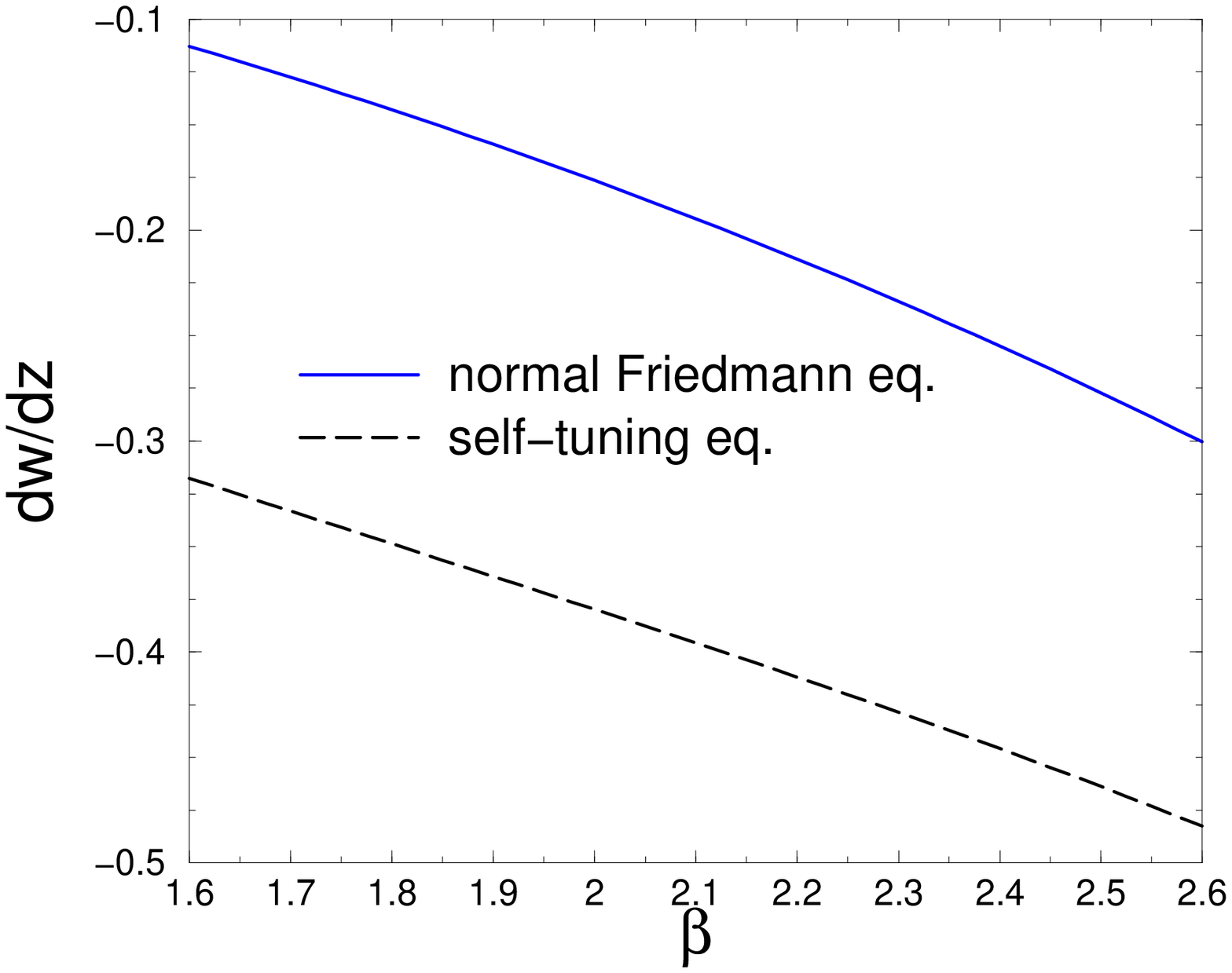}
\label{fig5}}
\vspace{-1.25cm}
\caption{Rate of change of the quintessence equation of state,
$dw/dz$, at $z=1/2$ as a function of the potential parameter $\beta$.
In the region of $\beta=2.4$, $dw/dz \approx 0.23 - 0.2\beta$
for the normal Friedmann equation, and near $\beta = 1.8$, 
$dw/dz \approx -0.07 - 0.16\beta$ for the self-tuning one.}}

\noindent {\bf Note added}: as this work was being finished, ref.\
\cite{halyo} appeared, which presents a different quintessence model that
also avoids the future horizon.

\acknowledgments

I thank the theory group of Lawrence Berkeley Laboratory for their
hospitality while this work was being finished.


\end{document}